# The Prediction and Interpretation of Singularities and Black Holes: From Einstein and Schwarzschild to Penrose and Wheeler

Dennis Lehmkuhl
Lichtenberg Group for History and Philosophy of Physics,
Institute of Philosophy, University of Bonn
Email: dennis.lehmkuhl@uni-bonn.de

*Abstract:* The Schwarzschild solution was the first exact solution to Einstein's 1915 field equations, found by Karl Schwarzschild as early as 1916. And yet, physicists, mathematicians and philosophers have struggled for decades with the interpretation of the Schwarzschild solution and the two singularities appearing in it when it is written in polar coordinates. This article distinguishes between eight different ways in which the two singularities have been interpreted between 1916 and the late 1960s, when Penrose's first singularity theorem shed new and lasting light on the interpretation of the Schwarzschild solution.

> "All bets were off concerning the possibility of a non-singular outcome to a collapse [of a star] after the theorem by Roger Penrose (1965), which has claims to be considered the most influential development in general relativity in the 50 years since Einstein founded the theory."
> —Werner Israel, 1987[1]

**Introduction**

In 2020 Sir Roger Penrose was awarded the Nobel Prize "for the discovery that black hole formation is a robust prediction of the general theory of relativity." Penrose shared the Nobel Prize with Rainer Genzel and Andrea Ghez, who together were awarded the other half of the 2020 Nobel Prize "for the discovery of a supermassive compact object at the center of our galaxy." [2] In May 2022, the Event Horizon Telescope Collaboration (EHT) published a picture of this supermassive compact object, which shows a dark, close to spherical object at the center and a sphere of light surrounding it. In the accompanying press release, the EHT stated: "Astronomers have unveiled the first image of the supermassive black hole at the center of our own Milky Way galaxy. This result provides overwhelming evidence that the object is indeed a black hole and yields valuable clues about the workings of such giants, which are thought to reside at the center of most galaxies." Likewise, in the corresponding scientific paper announcing the result, the EHT stated that "[w]e present the first Event Horizon Telescope (EHT) observations of Sagittarius A* (Sgr A*), the Galactic center source associated with a supermassive black hole."[3] It is striking that only two years earlier the Nobel Foundation felt comfortable justifying Penrose's half of the Nobel Prize for having derived black hole formation as a robust prediction of general relativity (GR from now on), and yet shied away from calling the object that Genzel and Ghez had meticulously observed in the center of the Milky Way a black hole.

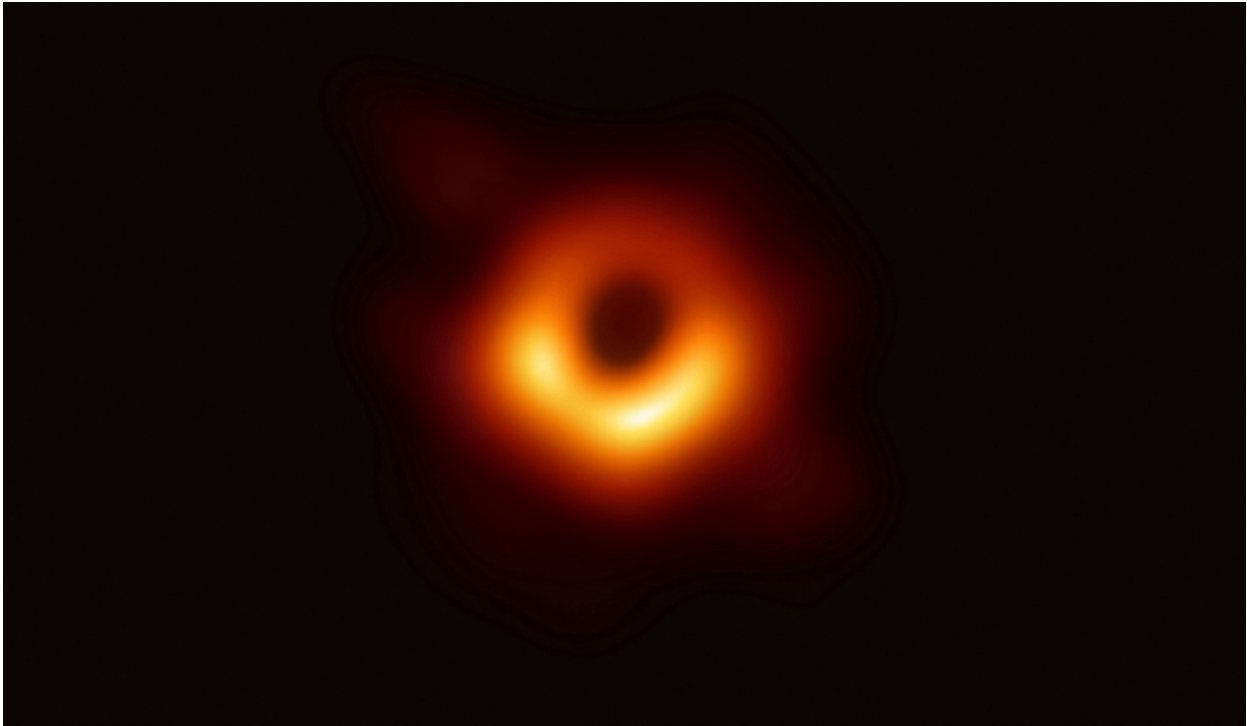

**Fig. 1:** The supermassive object at the center of the galaxy Messier 87, on April 10, 2019, announced as the first ever image of a black hole by the Event Horizon Telescope Collaboration. Credit: EHT Collaboration.

It is even more striking that in their more detailed justification of the prize, the Nobel Foundation explicitly referred to the 1965 paper in which Penrose proves the first singularity theorem,[4] and yet the term "black hole" does not appear in that paper, it only became widely used after a 1968 paper by John Archibald Wheeler.[5] Indeed, as Landsman has argued, one might well say that in Penrose's paper "the main differences between Penrose's actual theorem and its reputation are that: [a] The theorem says *nothing* about event horizons, which form the 'black' ingredient of a black hole; [b] It is *inconclusive* about 'singularities', which should form the 'hole' part of a black hole. Hence, the link between the theorem and black holes is opaque."[6]

This paper will address, first, the question of what spacetime singularities really are, and second how they came to be connected to black holes, through and in the aftermath of Penrose's 1965 theorem. But we can only really understand the significance of Penrose's theorem if we understand the soil from which it grew, i.e., only if we understand how physicists and mathematicians thought of singularities beforehand. Indeed, for a long time, both physicists and mathematicians puzzled not only about how spacetime singularities should be defined, but also about whether they could *actually* exist in nature, or whether they are instead artifacts of our mathematical reasoning.

I shall proceed in the following way. The following section, The Singularities of the Schwarzschild Solution, will discuss the first solution to the Einstein field equations ever found, the Schwarzschild solution, and how the fact that it seemed to feature two rather different types of singularities was interpreted first in the late 1910s and 1920s, and how Einstein's discussion with Jacques Hadamard at the Paris conference of 1923 shed new light on the issue, bringing physics within shouting distance of the concept of a black hole. This came about via the question of what would happen to a star if it were to be incredibly dense. I will introduce Eddington's 1920 model of how stars work, and Subrahmanyan Chandrasekhar's result that a very dense type of stars, white dwarves, cannot be denser than a certain threshold. I shall then discuss Oppenheimer and Snyder's pathbreaking paper of 1939, later cited in the introduction to Penrose's 1965 singularity paper, and how they argued that a sufficiently heavy star described by the Schwarzschild solution would collapse indefinitely and give rise to what would eventually be called a black hole. The third section, Singularities beyond the Schwarzschild Solution: Penrose's 1965 Theorem, will then discuss Penrose's path towards the singularity theorem of 1965, and how both the premises and the conclusion of the theorem fundamentally changed the way that the relativity community thought about singularities, and how they became associated with black holes. In the course of all this, I will distinguish between eight different interpretations of the term "singularity" that have been used by relativists through the ages and argue that even nowadays more than one of these interpretations is viable. The final section on The Introduction of "Black Hole" and "Cosmic Censorship" in the Aftermath of Penrose's 1965 Theorem will address how the concept of a black hole developed and how it is connected to Penrose's theorem.

**The Singularities of the Schwarzschild Solution**

*Finding the Schwarzschild Solution*

Our story starts in 1915, when Einstein introduced what we today call the Einstein field equations; the law of gravity that replaced Newton's law of gravity, and which was spectacularly confirmed in Arthur Eddington's observations of light bending by the Sun during the solar eclipse of 1919.[7]

The Einstein field equations read:[a]

---

[a] The left-hand side features the metric tensor $g_{\mu\nu}$, which plays the role of the gravitational potential in Einstein's theory but also gives the distance between any two points of spacetime; the Ricci tensor $R_{\mu\nu}$, itself the only non-trivial contraction of the Riemann tensor $R_{\mu\nu\sigma}^{\omega}$, which is defined in terms of first and second order derivatives of $g_{\mu\nu}$; and the Ricci scalar $R$, which is the contraction of the Ricci tensor in turn. The right-hand side consists of the coupling constant κ, which is essentially equivalent to Newton's gravitational constant $G$, and the stress-energy-momentum tensor $T_{\mu\nu}$, which encodes the mass and energy distribution of all material systems in a given spacetime. Einstein first wrote the field equations in this form in 1919. See: Albert Einstein, "Spielen Gravitationsfelder im Aufbau der materiellen Elementarteilchen eine Rolle?," *Sitzungsberichte der Königlich Preussischen Akademie der Wissenschaften* (1919), 349–56; reprinted in Michel Janssen, Robert Schulmann, József Illy, Christoph Lehner and Diana Kormos Buchwald, eds. *The Collected Papers of Albert Einstein* (Princeton: Princeton University Press, 2002), 7: 130–40 (Doc. 17); specific volumes and documents in these *Collected Papers* will be referred to hereafter as (CPAEx, Doc. y), where x is the volume and y is the document number. In their original form, the field equations had a term with the trace of the stress-energy-momentum tensor on the right-hand side instead of the term with the Ricci scalar on the right-hand side. See Albert Einstein, "Die Feldgleichungen der Gravitation." *Sitzungsberichte der Königlich Preussischen Akademie der Wissenschaften* (1915), 844–47; reprinted in CPAE6, Doc. 25. This version is mathematically equivalent to the version given here, which has become standard.

$$R_{\mu\nu} - \tfrac{1}{2} g_{\mu\nu} R = \kappa T_{\mu\nu} \tag{1}$$

Every solution to these equations represents a universe or a part of a universe that is *possible* according to GR, noting that this does not mean that it has to correspond to anything *actual* in our universe; for example, we are pretty sure that Gödel's solution has nothing whatsoever to do with the universe we actually live in. Solutions to the equations for which $T_{\mu\nu} \neq 0$ describe regions of spacetime in which matter is present; solutions for which $T_{\mu\nu} = 0$ correspond to vacuum regions of spacetime. In the latter case, the Einstein field equations reduce to the so-called vacuum Einstein equations,[8]

$$R_{\mu\nu} = 0. \tag{2}$$

In 1915, just a week before he published the gravitational field equations above, Einstein set out to describe a particularly important part of our universe. He wanted to derive the path of Mercury around the Sun. He had some hope that his new theory would shine compared to Newton's theory of gravity in this regard, for it had been known for a while that the prediction of Newton's theory was not quite in line with recent astronomical observations of Mercury. Furthermore, since Mercury is the planet closest to the Sun, it is the planet in our solar system that is subject to the strongest gravitational field, and thus most likely to uncover deviations from Newton's theory and Einstein's new theory, which contained Newton's theory as the limit of weak gravitational fields and slowly moving bodies.[9]

Einstein knew that he could treat Mercury itself as a test particle without a gravitational field of its own because its mass was so much smaller than that of the Sun. He also knew that he only needed to find a solution that could represent the Sun's *exterior* gravitational field that Mercury was subject to. Thus, he needed to find a solution to the vacuum Einstein equations (2). The fact that it was the exterior gravitational field of the Sun in particular Einstein incorporated by demanding that the gravitational field in question was to be a) spherically symmetric (as the Sun is an approximately spherically symmetric body); b) static (he assumed that the Sun does not change much, and so neither does its gravitational field); and c) asymptotically Minkowskian (meaning that far away from the Sun the gravitational field becomes so weak that it becomes indistinguishable from gravitation-free Minkowski spacetime).

These constraints already made the task of finding a solution much simpler. Still, Einstein believed that even so highly non-linear second order equations like (2) would be almost impossible to solve exactly. So he approximated the full vacuum Einstein equations by *linearizing* them, meaning that he only looked at the special case of weak and thus linear gravitational fields.[b] But an approximate solution of this kind was good enough to make a precise prediction of Mercury's anomalous perihelion, and to Einstein's excitement, the prediction corresponded precisely to the astronomically measured value.[10]

---

[b] Solutions of linear field equations, like the Newton-Poisson equation in Newton's theory of gravity or the Schrödinger equation in quantum mechanics, obey the superposition principle in a way similar to water waves: if solution *A* and solution *B* are solutions to the field equations, then their superposition *A* + *B* is also a solution. In the case of non-linear field equations this superposition principle fails, making the field equations much more difficult to solve. In addition, both Newton's and Schrödinger's equation only attribute a scalar field to every point (of space and configuration space, respectively), so only one field value per point, whereas the metric tensor $g_{\mu\nu}$ attributes ten field values to every point of spacetime.

Einstein was astonished when, only a few months later and writing from the trenches of World War I, Karl Schwarzschild informed Einstein that he had found the exact counterpart to Einstein's approximate solution: the first exact solution to the Einstein field equations and now immortalized as the Schwarzschild solution.[11] It fulfilled the same demands a) to c) as Einstein's approximate solution and could be assumed to describe the exterior gravitational field of spherically symmetric stars and planets of arbitrary size and mass.

As already hinted at, the metric tensor $g_{\mu\nu}$ that features in the Einstein equations gives rise to a so-called line element $ds^2$ that describes the distance between any two spacetime points. In the case of the Schwarzschild solution, this line element is (in polar coordinates)

$$ds^2 = -\left(1 - \frac{2m}{r}\right) dt^2 + \left(1 - \frac{2m}{r}\right)^{-1} dr^2 + r^2(d\theta^2 + sin^2\theta \, d\varphi^2), \qquad (3)$$

where *m* is interpreted as the mass of the Sun, or more generally the mass of the spherically symmetric body giving rise to the gravitational field in question. The interpretation of the parameter *m* as the mass of the gravitational source relies on the Newtonian limit of the Schwarzschild solution; one might even say that this interpretation "trickles up" or is imported from the Newtonian limit of the Schwarzschild solution.[12]

In the above coordinate system, the Schwarzschild solution exhibits two singularities. At the time "singularity" was typically meant to signify a point in space or spacetime at which the components of the gravitational potential $g_{\mu\nu}$ tend to infinity. The first singularity is at *r* = 0, i.e., at the center of the Sun and the coordinate system, where the first term of equation (3) tends to infinity. I will call this singularity the *central singularity*. The second singularity is at *r* = 2*m*, in which case the metric component in the second term of equation (3) tends to infinity. This singularity would soon be called "the Schwarzschild singularity" by Einstein and others; for ease of distinguishing it from the central singularity of the Schwarzschild solution I will instead call it the *spherical singularity*.[c]

*Einstein's and Schwarzschild's Interpretation of the Two Singularities in the Schwarzschild Solution*

The central singularity was not unexpected: the spherically symmetric solutions to Newton's gravitational equations have a singularity at their center too. Einstein and Schwarzschild interpreted it as a placeholder for the interior of the Sun that had not been included in the model.[13] This gives us the first of what will be seven interpretations of "singularity" that we shall encounter.

---

[c] The distinction between these two singularities is not as immediate as has sometimes been claimed in the literature. Schwarzschild himself restricted the coordinate system to $r > 2m$ and regarded $r = 2m$ as the center of his coordinate system, thus having "moved" the two singularities into one. The distinction did become common knowledge sometime during the early 1920s, as we shall see in the rest of the section. Different coordinate systems and different coordinate restrictions play a major role in this story; my colleague Christian Röken and I will soon publish a comprehensive technical paper on this.

The *placeholder interpretation*—Einstein's and Schwarzschild's interpretation of the central singularity of the Schwarzschild solution—essentially states that the singularity in question is nothing to worry about but something to be expected: it is an artifact of the fact that the source of the gravitational field has not been included in the modeling process. Thus, the singularity is a placeholder for a matter source, which is believed to be there but has not been included in the description yet. In the case of Einstein and Schwarzschild in 1916 the placeholder interpretation of the central singularity of the Schwarzschild solution is particularly plausible because only a solution to the vacuum field equations (2) had been sought and needed to calculate the influence of the Sun's gravitational field on Mercury.

> 1. Placeholder interpretation

During the early 1920s many relativists used what Eisenstaedt[14] has called the *pragmatic interpretation* but applied it specifically to the spherical singularity. This second interpretation likewise states that the singularity is a mathematical artifact, something that could not correspond to anything in nature, is not physically relevant. Thus, the pragmatic interpretation of the spherical singularity restricts the coordinates in the Schwarzschild metric to $r > 2m$. A paradigmatic statement to this effect was also given by Weyl in 1917: "Of course, only a segment of the solution that does not extend as far as the singular sphere can actually be realized in Nature."[15] Thus, the difference between the placeholder interpretation and the pragmatic interpretation is that the latter explicitly restricts the coordinates to get rid of the singularity whereas the former tolerates its presence.

> 2. Pragmatic interpretation

Schwarzschild did not leave it at modelling only the exterior gravitational field. In his second paper of 1916, he modelled the Sun itself as an incompressible and spherically symmetric fluid ball of constant overall mass density.[16] Schwarzschild managed to translate this model into a particular mass-energy tensor $T_{\mu\nu}$ that acts as a source term in the full Einstein equations (1) and solved the equations for the gravitational potential $g_{\mu\nu}$ both within and outside of the star. He also managed to match this interior solution for $g_{\mu\nu}$ to his previous exterior solution that modelled the gravitational field outside of the star. Most importantly, in this model, no central singularity appears for a star of the mass of the Sun; but Schwarzschild also predicted that for a star modelled this way, the pressure would become infinite at the center of the star if the radius is smaller than 9/8 times the Schwarzschild radius $r = 2m$.

*The 1922 Paris Conference and the Hadamard Catastrophe*

So nobody really worried about the central singularity of the (exterior) Schwarzschild solution; it had a counterpart in Newtonian physics. But the spherical singularity was soon seen as a real puzzle, despite the pragmatic way of excluding it from the maths by restricting the coordinates to $r > 2m$. Discussions of the nature of the spherical singularity came to a head in April 1922, when Einstein gave a series of lectures at the Collège de France in Paris. By then, Schwarzschild's solution in the coordinate representation (3) had become common knowledge, as had the fact that the solution thus represented featured the central singularity at $r = 0$ and the spherical singularity at $r = 2m$. At least, the Schwarzschild solution in this form features prominently—indeed is the only equation—in the report of Einstein's lectures for the *Revue des Deux Mondes*.[17]

At Einstein's lecture on 5 April 1922, Jacques Hadamard, one of the premier French mathematicians at the time, essentially posed the following question to Einstein: What if a star were so massive and so dense that its spherical singularity ended up *outside* of the star? For Hadamard knew that the Schwarzschild radius, the distance between the center of the star and the spherical singularity, was determined by the mass of the body, and for a particularly massive and small (thus dense) star could in principle be located outside of the star.

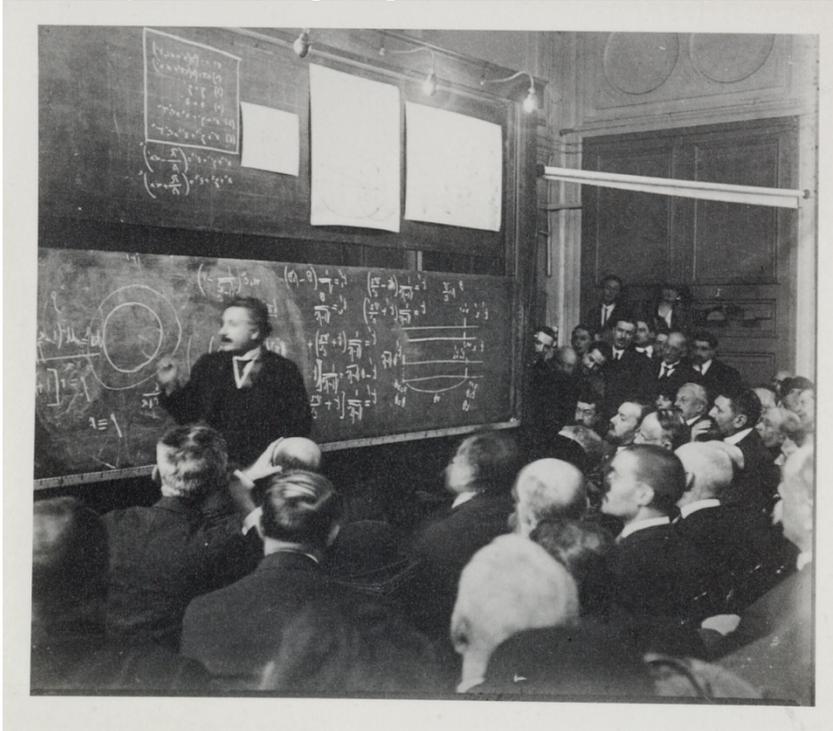

**Fig. 2**: Albert Einstein lecturing at the Collège de France in April 1922. Credit: Leo Baeck Institute.

Einstein did not mince words and answered that this would be a catastrophe for the theory. Two days later, on 7 April 1922, Einstein came back to the question and stated that "Hadamard's catastrophe" could never ever happen! He said that he calculated that before a star could accumulate sufficient mass to be entirely contained within its own Schwarzschild radius, the pressure at the center of the star would become infinite. As we have seen in the previous section, Schwarzschild made a similar prediction based on a model of the star that assumed perfect spherical symmetry and constant overall density. According to Charles Nordmann, Einstein now argued that if the pressure at the center of the star were to become infinite, then clocks would not run and thus no change could take place anymore, preventing the Hadamard catastrophe from ever developing.[d]

But why would it have been a catastrophe in the first place? The answer is that Einstein was fine with introducing singularities to *stand in* for matter: as long as a singularity could be interpreted as a placeholder for matter, as in his calculations of 1916, rather than be seen as something that actually exists, everything was fine for Einstein. That's why he did not worry about the central singularity in the Schwarzschild solution.

---

[d] Though the details of this argument are not carried out in Nordmann's report, we can easily see that a pressure singularity at $r = 0$ would bring about a singularity in the Einstein tensor $G_{\mu\nu} \equiv R_{\mu\nu} - \frac{1}{2} g_{\mu\nu} R$ via the Einstein equations (1). This can but must not in turn bring about a singularity in the metric tensor $g_{\mu\nu}$. If so, then changes in time would not be well-defined, though strictly speaking only very near to the singularity. In this sense, clocks would "stop."

But singularities were not supposed to actually *appear* in nature, and that's what was at stake in the debate with Hadamard: allowing for a star so massive and dense that its spherical singularity would be outside of the star would imply that a singularity would actually appear in empty space! Indeed, it is important to note that even though Einstein was fine with singularities acting as placeholders for a theory of matter not yet included in the model, just as he saw the energy-momentum tensor $T_{\mu\nu}$ in equation (1) as a temporary device to be used until a proper theory of matter was found, Einstein did believe that the vacuum equations (2) gave a complete and reliable theory of gravity, space and time. Thus, in the domain governed by the vacuum equations, the possibility of singularities appearing would have been a catastrophe, indeed arguably a falsification of GR, for Einstein.[18]

Discussion during the 1920s kept focusing on the nature of the spherical singularity, and how its appearance in equation (3) should be interpreted. Eddington wrote that it is "a magic circle which no measurement can bring us inside. It is not unnatural that we should picture something obstructing our closer approach and say that a particle of matter is filling up the interior."[19] Soon there was a real cottage-industry of coming up with ever new names for the spherical singularity, linked to different interpretational possibilities. Schwarzschild himself called it "the discontinuity," Friedrich Kottler "the barrier," Hermann Weyl sometimes "the sphere" and sometimes "the gravitational radius." Ernst Reichenbächer called it "the hole in the world,"[e] Marcel Brillouin "the singular frontier," "the catastrophic frontier" or "the limit circle," and Nordmann, who had reported on the 1922 Paris meeting, simply called it "the death."[20]

One can witness the emergence of an alternative interpretation of the spherical singularity in a variety of works during the 1920s, which I shall call the *barrier* (or *black shield*) *interpretation*: here the (spherical) singularity is interpreted as a kind of barrier through which a material system (like a particle or an observer) cannot pass. This barrier might be the surface of an extended material body, in which case the barrier interpretation is compatible with the placeholder interpretation.

> 3. Barrier interpretation

One example of this barrier interpretation might be seen in Eddington's quote above. But we have to be careful: this quote is from his semi-popular book from 1920. The context of the quote shows that he thought of the barrier as an *indicator* that there is a material source, or in the case of the spherical singularity the surface of a material source. Einstein had made a similar move when he interpreted the de Sitter solution of 1917. In his 1923 book directed at his peers from mathematics and physics instead of the general public, Eddington discussed both the singularities appearing in the de Sitter solution and in the Schwarzschild solution, struggling with what singularities may or may not indicate, writing: "A further question has been raised, is de Sitter's world really empty? In [the de Sitter metric] there is a singularity at $r = \sqrt{3/\lambda}$ similar to the singularity at $r = 2m$ in the solution for a particle of matter [i.e., the Schwarzschild solution]. Must we not suppose that the former singularity also indicates matter—a "mass-horizon" or ring of

---

[e] This is quite prophetic, given that, as we shall see below, the spherical singularity would eventually be found to be a coordinate singularity, and found to coincide, though only polar coordinates, with the event horizon of a black hole, a concept that only arose in the late 1950s. Indeed, it is remarkable that the man that happened to find what would come to be known as the first black hole solution happened to have the name Schwarzschild—German for "black shield." This is even more remarkable if one thinks about the fact (to be discussed below) that the event horizon of a black hole is a 3-sphere that one cannot pass again once one is within the black hole.

peripheral matter necessary in order to distend the empty region within?"[21]

This shows quite clearly that at least in the case of the Schwarzschild solution, and tentatively also in the case of the de Sitter solution, Eddington interpreted the singularity (in the case of Schwarzschild, the spherical one) as an *indicator*—a word he himself uses—for the presence of matter. Eddington, and Einstein before him, pioneered what is still a popular move in the interpretational analysis of exact solutions:[22] the *Indicator Interpretation*. Say we just found a new mathematical solution to the gravitational field equations, so a new shape the gravitational field can take. Assume that there is a singularity in the solution. This singularity may be seen as an indicator for the presence of matter sources that give rise to said gravitational field and are not yet modelled within the solution at hand.

<div style="float:right">4. Indicator interpretation</div>

Note that "being an indicator for something" is supposed to be different from "being a sufficient condition for something": it's a reason to entertain the *possibility* that there might be matter, whereas a sufficient condition would mean that matter *definitely* is there. If this possibility is substantiated by further investigations, for example by finding a corresponding solution to the full Einstein equations including the energy-momentum tensor $T_{\mu\nu}$ that can model this source, then the indicator of the vacuum solution turns into a placeholder for the non-vacuum source. This is a plausible way of interpreting Schwarzschild's two papers of 1916: the first paper gave the solution for the *exterior* (and thus vacuum) gravitational field of the Sun (or more generally of a spherically symmetric body), where the central singularity could have been seen as an *indicator* of the material source not yet modeled. The second paper took that indicator as a motivation to look for a solution that models the *interior* of the Sun, such that it would give rise to the exterior solution previously found.

The indicator and placeholder interpretations are closely related; the former focuses more on the context of discovery. It gives the relativist a reason to *look for* material sources, it encourages the search for an extended model including the material source. This is exactly what Schwarzschild did: in his second paper of 1916 he provided a solution for the interior gravitational field of the Sun and showed that the associated matter source fits with the previously found solution for the *exterior* gravitational field with its indicator singularity at the center. But now the solution representing the exterior gravitational field can be rid of its singularity by matching it with the interior solution and the associated (non-singular) matter source. In this way of seeing things, a singularity occurring in a vacuum solution to the Einstein equations is not so much a problem but a resource, something that tells you what to pursue next.

*The Interior of Stars*

We have seen in the previous section that the exterior Schwarzschild solution was initially interpreted, indeed designed, to represent the exterior gravitational field of stars. But no model of the star's interior was included in that description, and indeed one was not needed for the predictions that Einstein and Schwarzschild were after initially. However, we also saw that Schwarzschild quickly followed up on his paper on the *exterior* Schwarzschild metric with a second paper on what soon became called the *interior* Schwarzschild metric, which provides an explicit model for a star that gives rise to the exterior Schwarzschild metric. The model Schwarzschild provides in that second paper is that of an incompressible spherically symmetric fluid ball. Indeed, as we saw above, it is likely that this interior Schwarzschild solution inspired Einstein's arguments during the 1922 Paris meeting, predicting that Hadamard's catastrophe could not happen in reality.

However, Schwarzschild's model of a star, though an exact solution of the Einstein field equations, does not explain how stars actually work. In particular, it does not tell us why they *shine*, unlike planets or most of the stuff that we find around us on Earth. The first promising answer to this question was provided by Eddington in 1920. Eddington suggested that stars shine because of the energy produced in the nuclear fusion of hydrogen. The resulting outward-directed radiation pressure is what keeps the star from gravitationally collapsing.[23] But this immediately raised the question of what would happen when the star runs out of hydrogen to "burn." Then only the inward gravitational attraction would remain. Would the star collapse? And if so, would it find a new equilibrium state, become a smaller star—maybe less bright but still stable?

In 1930, Chandrasekhar and Lev Landau independently gave a definitive answer to this question for the main candidate of such a smaller star: a white dwarf. Chandrasekhar predicted that the mass limit for a stable white dwarf was 1.44 times the mass of the Sun; if a white dwarf was any heavier than this, then it would have to collapse further. With hindsight, Chandrasekhar's result already suggested that a sufficiently heavy star, when out of hydrogen, might just continue to collapse to ever smaller regions of space.[24]

This line of thought was further strengthened by a paper Robert Oppenheimer and Hartland Snyder published in 1939, entitled "On continued gravitational contraction."[25] Israel states that this paper "has strong claims to be considered the most daring and uncannily prophetic paper ever published in the field."[26] Oppenheimer and Snyder constructed an explicit, spherically symmetric model of a collapsing star whose exterior gravitational field was given by the exterior Schwarzschild solution, and whose interior was modelled by a collapsing fluid ball whose internal pressure could be neglected and had run out of "nuclear sources of energy".[27] They predicted that a sufficiently heavy star of this kind would continue to collapse without ever achieving a new equilibrium state, collapse to within the sphere formed by its spherical singularity. They also predicted that once this happened, "an observer comoving with the matter would not be able to send a light signal from the star; the cone within which a signal can escape has closed entirely."[28] This is the first appearance of the idea that a sufficiently heavy collapsed star would trap light, and everything else inside it—one reason why such a collapsed star would later be named a "black hole" (see the final section of this paper). Oppenheimer and Snyder predicted that, as a result of this collapse, the central singularity of the Schwarzschild solution would form. Thus, if Oppenheimer and Snyder are correct, then what Einstein and Schwarzschild thought of as a placeholder for matter could actually come about *in nature*.

So did Oppenheimer and Snyder falsify Einstein's reasoning of April 1922? I think not: they make the idealizing assumption that the pressure inside the star vanishes, whereas for Einstein the assumption of non-vanishing pressure (in line with Schwarzschild's interior solution) played a major role in his argument. Indeed, Oppenheimer's and Snyder's analysis rests on two very specific assumptions: i) that the gravitational field of the collapsing star is exactly spherically symmetric, and ii) that the matter that the star is made of can be idealized as pressureless dust.[29]

Oppenheimer and Snyder's paper went almost unnoticed for a number of years, but it had a huge comeback during the 1950s and 1960s. For, as we shall see in the following section, in these two decades, now often heralded as a renaissance of GR, the question of continued gravitational collapse unexpectedly connected both to new astronomical observations and to new mathematical techniques.

*Wheeler Reconsidering the Central Singularity*

John Wheeler was a major reason for the fact that Oppenheimer and Snyder's paper finally came to prominence during the 1950s. Wheeler's main background was in nuclear physics, but in the early 1950s he shifted his research interests to GR. Oppenheimer and Snyder's paper was the perfect overlap between the two fields. And maybe it was exactly for that reason that Wheeler was so unhappy with the paper, for he felt that the assumptions made for modelling the interior of the collapsing star could not possibly come close to doing justice to an *actual* star, and that thus any consequences derived from that star model (like that of the emergence of a singularity in the center of the star) should be met with extreme caution. In 1958, Wheeler, Harrison and Wakano reconsidered the Oppenheimer-Snyder collapse model, using some of the earliest computers to model the gravitational collapse of a star differently and more in line with nuclear physics.[30] In the same year, there was a head-on collision between Wheeler and Oppenheimer at the 1958 Solvay conference in Brussels. Wheeler reported on his work with Harrison and Wakano, and put their conclusions as to what would happen to a star in the final stages of its collapse as follows: "Perhaps there is no final equilibrium state: this is the proposal of Oppenheimer and Snyder … A new look at this proposal today suggests that it does not give an acceptable answer to the fate of a system of A-nucleons under gravitational forces."[31] Oppenheimer was in the audience and asked: "Would not the simplest assumption about the fate of a star more than the critical mass be this, that it undergoes continued gravitational contraction and cuts itself off from the rest of the universe?"

Wheeler was unconvinced by Oppenheimer and remained so way into the 1960s. He expected that a proper model of the collapsing star would have to take into account the repulsive forces between the nucleons making up the star and for that a proper quantum-theoretic description of the star—and most likely a quantum theoretic description of the gravitational forces between the nucleons—would be necessary in these interior regions where matter was incredibly dense and gravity incredibly strong.

Thus, I believe that Wheeler's interpretation of the central singularity in the Schwarzschild solution, as it appears in the Oppenheimer-Snyder collapse model, is best termed the *problem interpretation*. According to this interpretation the presence of a singularity in the solution is the sign of a problem: the model or theory is inadequate or at least incomplete. (Indeed, note that Einstein's reaction to what he termed "Hadamard's catastrophe," and which he considered a counterfactual, can be seen as endorsing a version of the problem interpretation.) The model that brings about such a singularity needs to be replaced by something better. In principle, this replacement could be achieved by filling in a placeholder, but if we take Wheeler as a paradigmatic problem interpreter, then it seems clear that he considered the problem to be likely too big for it to go away in this way. One could also call the interpretation a *breakdown interpretation*: the presence of a singularity is a signifier of a fundamental problem, of a breakdown of the model/theory employed, and thus a cry for new physics. It is then a judgement call whether a new model from within GR is expected to suffice for solving the problem or whether a new fundamental theory is required. This interpretation, too, is still very prominent today, with many of those working on a quantum theory of gravity using the problem/breakdown interpretation of singularities as an argument for why a quantum theory of gravity is urgently needed.

| 5. Problem interpretation |

*Finkelstein Reconsidering the Spherical Singularity*

Since the beginning of the 1920s, relativists had analyzed the Schwarzschild solution by using ever new coordinate systems. Some of them got rid of the puzzling and now famous "Schwarzschild singularity," which I called the spherical singularity, and in these coordinate systems it looked like a completely normal part of spacetime. Arthur Eddington and George Lemaître in the 1920s, and H.P. Robertson in the 1930s, clearly found such regularizing coordinate systems. But the interpretation of these coordinate systems remained controversial, and knowledge of them did not really catch on in the then quite dispersed relativity community. Even when Finkelstein discovered a coordinate system similar to Eddington's in 1958, many still spoke of the "Schwarzschild singularity." Indeed, in this period Finkelstein was the first who clearly said: "The Schwarzschild surface $r = 2m$ is not a singularity but acts as a perfect unidirectional membrane: causal influences can cross it but only in one direction."[32] What Finkelstein realized was *not only* that the spherical singularity is not really a singularity, he also realized that *it is something else*, something rather specific. He called it a unidirectional membrane, and following Rindler's 1956 clarification of the concept of an event horizon, the idea of a unidirectional membrane was soon absorbed into the concept of an event horizon.

Finkelstein's path towards this way of thinking was brought about by a) his finding a coordinate system similar to the one Eddington had obtained where $r = 2m$ is not singular; b) focusing on the behavior of null rays (paths of light) crossing $r = 2m$; c) thinking about the behavior of the Schwarzschild solution under time reversal transformations $t \rightarrow -t$ in his coordinate representation. This last point contains the seed of the later distinction between black holes on the one hand and white holes on the other.

Finkelstein's interpretation of the spherical "singularity" of the Schwarzschild solution could be called the *transformative* (or *black membrane*) *interpretation*: the spherical "singularity" of the Schwarzschild solution was actually something else in disguise: a membrane that can be passed from one side of the sphere but not from the other. As we shall see in the subsection on Penrose's Path to Relativity below, Finkelstein's realization was of absolutely crucial importance for Penrose's singularity theorem; and indeed for Penrose working on GR in the first place.

| 6. Transformative interpretation |

*The 1963 Texas Conference of Relativistic Astrophysics*

In the early 1960s, the Russian school of GR too took aim at the central singularity of the Schwarzschild solution, and especially at Oppenheimer and Snyder's claim that such a singularity would form during the gravitational collapse of a sufficiently heavy star. Evgeny Lifshitz and Isaak Khalatnikov, based on previous work by Yakov Zel'dovich and Igor Novikov, argued that Oppenheimer and Snyder's analysis of gravitational collapse was, as Penrose later put it, "not representative of the general situation,"[33] i.e., not representative for a star for which even small deviations from spherical symmetry were allowed. Indeed, Lifshitz and Khalatnikov had argued that a generic solution to the vacuum Einstein equations involves eight arbitrary functions whereas a singular solution involves only seven; and that thus singularities can't be a generic feature of solutions to the vacuum Einstein equations.[34]

That in turn suggested that Oppenheimer and Snyder's prediction that a central singularity would form in a collapsing star was a mathematical artifact produced by their assumption of perfect spherical symmetry of said star.

In a way, all this meant that, though strange, the spherical "singularity" was not as big a problem as physicists had long thought it to be; Finkelstein's work had finally made it known that it was only an artifact of the polar coordinate system. In contrast, the central singularity of the Schwarzschild solution, which originally nobody was worried about, had become more and more threatening with Oppenheimer and Snyder's analysis of gravitational collapse. For there it looked as if such a singularity could *actually* arise in nature, as the result of a *physical* process— the very thing that Einstein had termed a "catastrophe." And yet, Wheeler, Zeldovich, Lifshitz and Khalatnikov had argued that it too was a mathematical artifact, in this case a result not of a particular coordinate system but of the assumption of spherical symmetry that Oppenheimer and Snyder relied on. (In the next section, we shall see that Penrose's theorem can be seen as a direct counter to the line of argument championed by Wheeler, Zeldovich, Lifshitz and Khalatnikov.)

Everything I've said so far was about theoretical GR and how gravitational collapse of a star could *in principle* happen. There was, thus far, no relationship to any actual astronomical bodies. This all changed in 1962. Maarten Schmidt, after receiving data from Cyril Hazard, M.B. Mackey and A. J. Shimmins, announced the observational discovery of a new type of quasi-stellar object, or quasar for short.[35] This discovery was the impetus for organizing the first Texas Conference of Relativistic Astrophysics in 1963, which brought together mathematical relativists and observational astrophysicists for the first time. It was soon clear that a quasar can emit more energy than our entire Milky Way Galaxy. Now one might ask how it could be that these objects were only discovered in 1962, if they are so energetic. The reason is that by far most of the energy is emitted in the X-ray and radio spectrum, and we just can't see X-rays and radio waves with our bare eyes or indeed with our optical telescopes. So X-ray and radio telescopes had to be developed before quasars could be observed.

How is all this related? The idea quickly came up that the source of all the energy emitted by a quasar could be the gravitational collapse of a very heavy star; the death woes of a star producing an enormous final burst of energy. Fred Hoyle and William Fowler wrote: "Our present opinion is that only through the contraction of a mass of $10^7$–$10^8 M_\odot$ to the relativity limit can the energies of the strongest sources be obtained."[36] So suddenly all the calculations by mathematical relativists about gravitational collapse became astutely relevant to astrophysics. Indeed, even the connection to singularities was made in the invitation to the Texas Conference that was sent out by Ivor Robinson, Alfred Schild, Engelbert Schücking and Peter Bergmann. In the invitation they wrote that the discovery of quasars and the theory put forward by Hoyle and Fowler "open up the discussion of a wealth of exciting questions," among them "(c) Does gravitational collapse lead, on our present assumptions, to indefinite contraction and a singularity in spacetime?"[37]

Roger Penrose was present at this conference and must have been on high alert given that the very invitation to the conference posed the question that had made him enter the field of GR in the first place five years earlier.

**Singularities Beyond the Schwarzschild Solution: Penrose's 1965 Theorem**

*Penrose's Path to Relativity*

Roger Penrose completed his PhD in pure mathematics on "Tensor methods in algebraic geometry" at the University of Cambridge in 1958. Already as a mathematics student he flirted with modern physics, specifically by attending the quantum mechanics lectures given by Paul Dirac and Hermann Bondi's general relativity lectures, and, maybe most importantly, by regularly interacting with the cosmologist Dennis Sciama. It was because of Sciama's advice that Penrose attended a lecture by David Finkelstein at King's College London in 1958, in which Finkelstein spoke about the Schwarzschild solution of the Einstein field equations. Finkelstein was giving the paper we discussed above and showed that the spherical singularity in Schwarzschild coordinates can be transformed away by changing the coordinate system in which the Schwarzschild solution is represented. And he showed that $r = 2m$ in Schwarzschild's coordinates was the location of what he called a "unidirectional membrane"—soon to be renamed an event horizon, and not an artifact of the coordinate system!

After attending this talk, Penrose shifted his focus of work to GR. In an interview with Alan Lightman on 24 January 1989, Penrose described what a profound influence Finkelstein's talk had on him: "I remember being struck by the fact that although you remove the apparent singularity at $r = 2m$ (the horizon of the black hole), you still had another singularity at $r = 0$ [the central singularity]. It seemed that you just pushed the problem somewhere else. When I got back to Cambridge, knowing very little about general relativity, I started to try and prove that singularities were inevitable. It seemed to me that maybe this was a general feature—that you couldn't get rid of the singularity. I really had no means of proving this at the time."[38]

It would take six years until Penrose finally proved a corresponding theorem, in the fall of 1964, a few months after the first Texas Conference that had refocused his mind on the problem he had set himself in 1958. In the meantime, he had developed quite a few new mathematical concepts and tools that went into the eventual theorem.

*The Premises of Penrose's 1965 Theorem*

Having witnessed the struggle with the interpretation of the spherically symmetric Schwarzschild solution between 1916 and 1958, we are now in a position to better understand the project that Penrose outlines at the beginning of his 1965 paper, entitled "Gravitational Collapse and Space-Time Singularities."

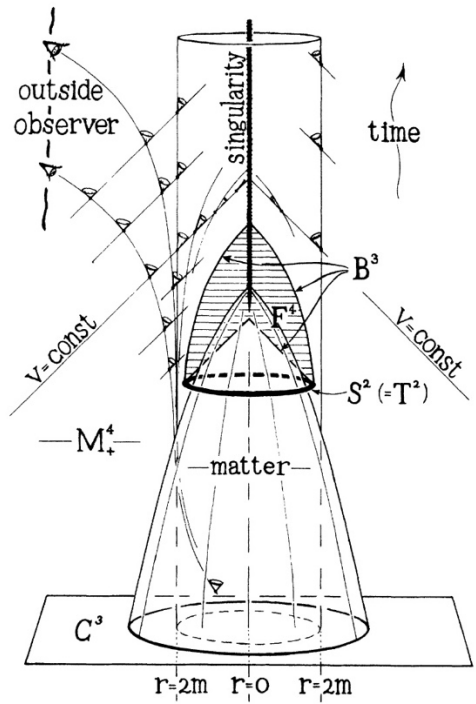 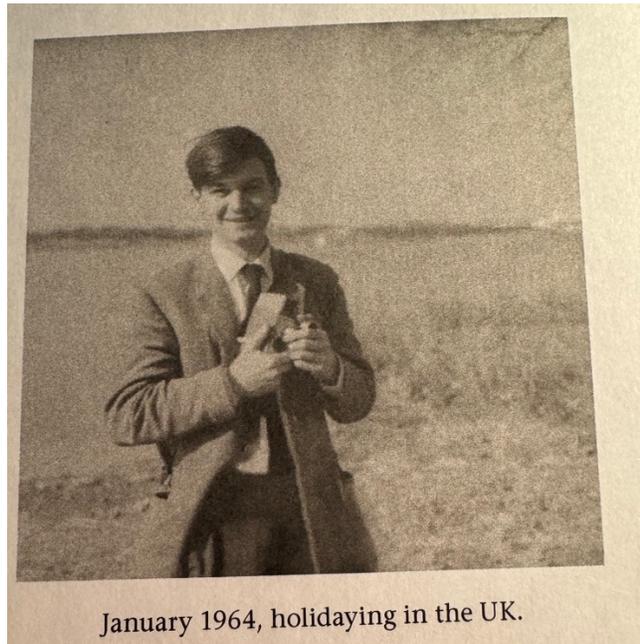

January 1964, holidaying in the UK.

**Fig. 3a**: Penrose's drawing of a spherically symmetric collapsing star as published in his 1965 paper. The Cauchy surface $C^3$ is a 3-dimensional spatial slice of spacetime at the time when the collapse of the star, represented by the outer circle on $C^3$, begins. The star keeps shrinking until it collapses to within the trapped surface $T^2$ at r=2m, which contains the central singularity at r=0. Once inside the trapped surface, all future light cones point at the singularity, representing the conclusion of Penrose's theorem that the corresponding geodesics will all end up incomplete because of hitting the singularity. Source: Roger Penrose, "Gravitational Collapse and Space-Time Singularities," *Physical Review Letters* **14**, no. 3 (1965), 57–59. **Fig. 3b**: Roger Penrose in 1964. Credit: Roger Penrose, Collected Works, Volume 1, front page.

    In the introduction, Penrose writes: "[M]ost *exact* calculations concerned with the implications of gravitational collapse have employed the simplifying assumption of spherical symmetry … The general situation with regard to a spherically symmetric body is well known … When sufficient thermal energy has been radiated away, the body contracts and continues to contract until a *physical singularity* is encountered at [the center of the body]. As measured by local comoving observers, the body passes within its Schwarzschild radius … The question has been raised whether this singularity is, in fact, simply a property of the high symmetry assumed. It will be shown that … deviations from spherical symmetry cannot prevent space- time singularities from arising."[39] Given our discussion above, we understand what Penrose refers to when he says that "most exact calculations concerned with the implications of gravitational collapse have employed the simplifying assumption of spherical symmetry"; here he speaks of the calculations of Oppenheimer and Snyder, just as when he claims that in these calculations "the body contracts and continues to contract until a *physical singularity* is encountered at [the center of the body]." Indeed, a footnote to "The general situation with regard to a spherically symmetric body is well known" contains references to Oppenheimer and Snyder on the one hand and Wheeler on the other, both discussed above. We also now understand what Penrose means with the term

"physical singularity"; it is in contrast to the unphysical coordinate singularity at $r = 2m$ of the Schwarzschild solution that Finkelstein had unmasked in front of Penrose's eyes. And we can also see clearly that a proof that showed that Oppenheimer and Snyder's result of a physical singularity arising at the center of the star does not depend on spherical symmetry would be a game-changer: it would be just as bad as what Einstein had called "Hadamard's catastrophe," the prediction of an actual singularity in nature, and one that could not be blamed on idealizing assumptions. Penrose was consciously arguing against Wheeler, Lifshitz and Khalatnikov on this, all of whom he cites; facing an unlikely alliance of big shot physicists from both the USA and the Soviet Union, in the middle of the Cold War and maybe thus all the more powerful.

And for Penrose, the paper was the culmination of the very idea that had made him work on GR, which matured into the result announced in the introduction of the paper, namely that "deviations from spherical symmetry cannot prevent space-time singularities from arising."

Let us first look at the structure of the theorem that led Penrose to this conclusion. I here give it in a form that is different from Penrose's original paper.[40] It states

> If a time-orientable spacetime fulfills
> - Premise 1: An energy/curvature condition: local energy is non-negative;
> - Premise 2: A causality condition: the spacetime has an initial Cauchy hypersurface;
> - Premise 3: An appropriate initial and/or boundary condition: there exists a trapped surface in the spacetime;
> 
> then it contains incomplete geodesics, and thus the spacetime is singular.[f]

Let us look at Premise 1 first. By far most previous treatments of gravitational collapse had assumed a very specific model for the collapsing matter; in contrast Penrose only assumed that whatever the collapsing star is made of, and no matter whether it is subject to pressure or not, its local energy is not negative. Thus, compared to what had often been assumed before, Premise 1 is an incredibly minimal assumption about the matter concerned.[41]

Now to Premise 2. Penrose defines a Cauchy surface by the property that every inextendible timelike or null curve (and thus every particle and every light ray) meets said surface, and that every curve intersects the surface exactly once; thus, together with the gravitational and matter field equations, ensuring the possibility of deterministic predictions given an initial such Cauchy hypersurface.[42]

But it was Premise 3, involving the concept of a trapped surface, that was truly revolutionary. For it was one of the first *quasi*-local, rather than purely local, concepts in GR, and it managed to skirt around all the previous problems involving coordinate systems and symmetries.

---

[f] In "Gravitational Collapse and Space-Time Singularities" (ref. 4), Penrose puts it slightly differently by arguing that what I call Premises 1 to 3 and the conclusion that the spacetime is geodesically complete are four mutually inconsistent statements. He then implicitly states that the statement that the spacetime is geodesically complete is the only one that can be wrong for the spacetime in question, and that thus, given the assumed truth of the first three statements, the spacetime must be geodesically incomplete. There are no arguments in this first paper for why the inconsistency between the four statements should be blamed on the fourth statement; I will have to leave tracing Penrose's later steps on this (as well as his evolving thought on the proof of the theorem) for a future paper.

Premise 1 translated into a condition on the Einstein tensor (via the Einstein field equations), Premise 1 and Premise 3 together imply a *focusing* of all infalling light towards the interior once the light rays are within the trapping surface.

It is today hard to fully appreciate the ingenuity that the concept of a trapped surface demanded at the time. Note that it was inspired by a coordinate system (that of Eddington-Finkelstein), and yet the concept itself does not need any reference to a coordinate system, and it does not rely on the spacetime having any symmetries. It boldly allows for a quasi-local property of spacetime, that of spacetime possessing a trapped surface, while the consensus for half a century had been that only local concepts are to be trusted. That had been the whole push towards GR, away from Newtonian theory! So here is the basic idea of a trapped surface: after the star has collapsed to within its Schwarzschild radius (which has now been understood to not be a genuine singularity), a sphere can be found in the empty regions surrounding the collapsing matter. The sphere is an example of a trapped surface: light rays emitted from within the sphere, no matter in which direction, will always converge inwards. The formal definition of a trapped surface Penrose introduces in 1965 is that it is a "closed, spacelike, two-surface $T^2$ with the property that the two systems of null geodesics which meet $T^2$ orthogonally converge locally in future directions at $T^2$."[43] Once Penrose had the concept of a trapped surface, it must have been immediately clear to him that the Schwarzschild solution, the Reissner-Nordström solution as well as the recently discovered Kerr solution[44] featured such trapped surfaces.[g] One of the immediate consequences of Penrose's theorem was that the central singularity of the Schwarzschild solution was not an artifact of the solution's spherical symmetry; for any solution obeying premises 1 to 3 would be geodesically incomplete, like the Schwarzschild solution is at $r = 0$.[h]

The first two premises were so general that one might well have argued that they delineate a big if not the entire subset of all physically relevant solutions of the Einstein equations. So it was the third premise concerning the existence of a trapped surface in a given spacetime that really singled out the singular nature of the solutions that Penrose's theorem was concerned with, and thus he put the result of the paper thus: "[T]he existence of a trapped surface implies—irrespective of symmetry—that singularities necessarily develop."[45]

So how did Penrose conceive of the idea of a trapped surface? In an interview with Kip Thorne in 1994 the story was first recorded, though he has since recounted it many times, especially after being awarded the Nobel Prize, in words very similar to those he chose when speaking to Thorne:

> My conversation with [Ivor] Robinson stopped momentarily as we crossed a side road, and resumed again at the other side. Evidently, during those few moments an idea occurred to me, but then the ensuing conversation blotted it from my mind! Later in the day, after Robinson had left, I returned to my office. I remember having an odd feeling of elation that I could not account for. I began going through in my mind all the various things that had happened to me during the day, in an attempt to find what it was that had caused this elation. After eliminating numerous inadequate possibilities, I finally brought to mind the thought that I had had while crossing the street.[46]

---

[g] As already noted, the Schwarzschild solution could be used to represent the exterior gravitational field of a spherically symmetric, static and electrically uncharged star; the Reissner-Nordström solution would be fit to represent such a star that also has electric charge; and the Kerr solution a rotating stationary star without electric charge.

[h] Penrose's theorem applies to the entire family of Schwarzschild spacetimes, but it is less clear to what subset of the families of Reissner-Nordström spacetimes and Kerr spacetimes, respectively, it applies. For even though both of them feature trapped surfaces, they also feature so-called Cauchy horizons, and the extent to which these can, for certain subsets of the family, get in the way of Premise 2, is a tricky question.

This sounds a bit as if the idea of a trapped surface just fell from the sky. But as so often in these cases it was more likely the culmination of seven years of pondering, and of developing puzzle pieces in the form of new mathematical concepts that simply fell into place when Penrose crossed the road with Robinson.[i] Let us now look at this in more detail.

After Penrose had attended David Finkelstein's talk on 28 January 1958 and decided to start working on GR and on proving that one can't get rid of the central singularity in the Schwarzschild solution the way Finkelstein had with the spherical singularity, Penrose started to work on recasting GR into a new form. Since his mathematical background was in tensor calculus, and since he had been inspired by Dirac's lectures on spinors, he set about reformulating GR in spinor form. He gave his first talk on his results at the Royamount Conference in June 1958, after Sciama had given up part of his allotted speaking time to make room for Penrose.[47] A more comprehensive paper on "A Spinor Approach to General Relativity" came out in *Annals of Physics* in 1960, a full two years before the earlier Royamount paper came out in the conference proceedings. In these two papers, Penrose rederived many core results of GR in a significantly simplified way, aided by the new spinorial form. From 1959 to 1961 he was a NATO Research Fellow, allowing him extended research stays in Princeton (with Wheeler's group), Syracruse (in Bergmann's group) and Cornell. Among other things he worked out "the larger portion" of a paper that would only be published in 1965, entitled "A Remarkable Property of Plane Waves in General Relativity."[48] The "remarkable property" that Penrose found was that such a gravitational wave spacetime does not possess a spacelike Cauchy hypersurface; and it was in this context that Penrose worked out his concept of what it is for a spacetime to have a Cauchy surface, which he would use in Premise 2 of the 1965 singularity theorem. Also while at Princeton, in 1960–61 (and again it took until 1966 for the paper to come out), Penrose worked out the idea that in a certain limit Ricci curvature and Weyl curvature could focus light analogously to purely anastigmatic and purely astigmatic lenses, respectively.[49] This would essentially give him Premise 1 of the 1965 theorem: the idea that Ricci curvature, as one would have it in the environment of a collapsing star, can focus light inward. Finally, also while at Princeton, Penrose started his lifelong collaboration with Ezra (Ted) Newman at the University of Pittsburgh. They combined their previously separate spinorial and tetrad formalisms in what was soon to be called the Newman-Penrose formalism;[50] and it was this tool/framework that Penrose used in the *proof* of his 1965 theorem three years later. In the following years Penrose developed his "conformal treatment of infinity" and the corresponding Penrose diagrams, and in the summer of 1964 he finished "Zero Rest-mass Fields Including Gravitation: Asymptotic Behavior."[51] The appendix of this paper developed techniques that would be of crucial importance for the proof of his singularity theorem in December 1964.

---

[i] I can only hint at the most superficial reconstruction of this in the following paragraph; I hope to properly disentangle Penrose's precise route together with colleagues in the next years, and we will publish a much more technical paper then.

All that had been missing was the idea of a trapped surface, Premise 3 above … and between crossing the road on that day and finding the proof of his first singularity theorem it did not take long. Let us now turn to the conclusion of the theorem, and to the question of what the kind of concept of singularities introduced there has to do with black holes.

*The Conclusion of Penrose's 1965 Theorem*

We have seen that previously whenever most physicists talked about a singularity, what they referred to was that the components of the metric tensor or the curvature tensor would tend towards infinity. The problem was that you could never really be sure whether this behavior was due to of the coordinate system chosen or whether it reflected something deeper.[j] The idea of moving away from thinking about whether the metric or the curvature components become infinite and towards thinking about whether the path of a light ray is complete or instead suddenly stops (which is the incomplete geodesic referred to in Penrose's conclusion), and to take this as a criterion for there to be a singularity, was another game-changer. Like his use of an energy condition and of the condition of the spacetime having a spacelike Cauchy hypersurface, it was Penrose who put these ideas to unprecedented use; but the idea itself was already present in the literature before him.

Among Penrose's predecessors, it was Charles Misner who most clearly brought out a relationship between geodesic incompleteness and the definition of a singular spacetime, suggesting that the former was a necessary but not sufficient condition for the latter.[52]

Now, here is again Penrose's conclusion from the 1965 paper: "[T]he existence of a trapped surface implies—irrespective of symmetry—that singularities necessarily develop."[53] As noted, the singular behavior that he speaks of here is that the path of a light ray cannot be extended further but instead stops: the light ray in question moves on an incomplete geodesic. In the case of the Schwarzschild spacetime, this happens at the center of the collapsing star, and the outermost trapped surface is located at the Schwarzschild radius, i.e., at the event horizon.

But what do incomplete geodesics have to do with singularities? In his 1965 paper, Penrose points out that the central singularity in the Schwarzschild solution is a "physical singularity", and commenting on his 1965 theorem four years later, he writes: "[We] know … that there must be some space-time singularity resulting inside the collapse region. However, we do not know anything about the detailed nature of this singularity."[54]

This does not sound as if Penrose just *identified* geodesic incompleteness with there being a singularity. Rather, it seems that he took geodesic incompleteness as an indicator for there to be a singularity, that which causes the geodesic incompleteness. The singularity is the *reason* for the geodesic to "stop"—it ran into the singularity and thus was rendered incomplete. Thus, I think Penrose's interpretation of what it is to be a singularity, at least in the late 1960s, is a mixture of the following two interpretations. First, I believe Penrose implicitly endorsed what one might call the *Indicator Interpretation 2.0*, the idea that geodesic incompleteness is an indicator for the presence of a physical, i.e. semi-localizable, singularity.[55] Second, Penrose is naturally interpreted as endorsing what one might call the *Barrier Interpretation 2.0*, defined as the idea that a physical singularity is a kind of barrier that brings geodesics to a premature end

| 7. Indicator interpretation 2.0. |
| 8. Barrier interpretation 2.0. |

---

[j] This claim is true about a singularity occurring in any of the metric tensor's $g_{\mu\nu}$ or the curvature tensor's $R_{\mu\nu\sigma}^{\omega}$ components. However, a singularity in the Kretschmann curvature scalar $K = R_{\mu\nu\sigma\rho}R^{\mu\nu\sigma\rho}$ is independent of the choice of the coordinate system.

(i.e., it *makes* the geodesics incomplete).q

Penrose's stronger demands for calling something a spacetime featuring a physical singularity, which demanded more than the spacetime being geodesically incomplete, is in contrast to Stephen Hawking, who first wrote in his PhD thesis that "any model must have a singularity, that is, it cannot be a geodesically complete $C^1$, piecewise $C^2$ manifold."[56] Much of the relativity community soon followed Hawking on this, though everyone accepted that there is a difference between a singular spacetime in Hawking's sense and the occurrence of a semi-localizable singularity like that at the center of the Schwarzschild solution.

**The Introduction of "Black Hole" and "Cosmic Censorship" in the Aftermath of Penrose's 1965 Theorem**

I have noted at the beginning that the term "black hole" was only introduced by Wheeler in 1968, three years after the paper that won the Nobel prize for predicting the existence of black holes. I also noted that Landsman argued that the 1965 paper did indeed need a further ingredient to establish the existence of black holes. That ingredient, he argued, is Penrose's later concept of cosmic censorship.

Let us start with Wheeler's coinage of the term "black hole." In his quasi-autobiography, he recalls:

> In the fall of 1967, [I was invited] to a conference … on pulsars. … In my talk, I argued that we should consider the possibility that the center of a pulsar is a gravitationally completely collapsed object. I remarked that one couldn't keep saying "gravitationally completely collapsed object" over and over. One needed a shorter descriptive phrase. "How about black hole?" asked someone in the audience. I had been searching for the right term for months, mulling it over in bed, in the bathtub, in my car, whenever I had quiet moments. Suddenly this name seemed exactly right. When I gave a more formal Sigma Xi-Phi Beta Kappa lecture … on December 29, 1967, I used the term, and then included it in the written version of the lecture published in the spring of 1968.[57]

In the written version of the lecture mentioned, Wheeler took full advantage of his skills as a writer when introducing the term. In the section entitled "The Black Hole," he writes:

> If gravitational collapse is as inescapable in a star as geometrodynamics is inescapable in the universe, then what would be the appearance of the collapsing core if it could be seen from afar ...? The hot core material is brilliant and at first it shines strongly into the telescope of the observer. However, by reason of its faster and faster infall it moves away from the observer more and more rapidly. The light is shifted to the red. It becomes dimmer millisecond by millisecond, and in less than a second too dark to see. What was once the core of a star is no longer visible. The core like the Cheshire cat fades from view. One leaves behind only its grin, the other, only its gravitational attraction. Gravitational attraction, yes; light, no. No more than light do any particles emerge. Moreover, light and particles incident from outside emerge and go down the black hole only to add to its mass and increase its gravitational attraction.[58]

Wheeler's talk took place two years after the publication of Penrose's theorem; how did it influence his view of what had once been a "gravitationally completely collapsed object" and was now a "black hole"?[59] Did Wheeler now believe, contrary to his earlier opposition against Oppenheimer and Snyder, that at the center of a black hole there is a spacetime singularity?

In the same paper in which Wheeler likens a black hole to the Cheshire cat, he comments on Penrose's and Hawking's singularity theorems: "No one knows a cheap way out. The infinity is an infinity so long as one stays within the context of classical theory. Infinity is a signal that an important physical effect has been left out of account. The root of the new physics is not far to seek. It is the quantum of action."[60] So even after Penrose's theorem, Wheeler still subscribed to the problem interpretation: what Penrose's theorem did in Wheeler's mind was only to reinforce the need for a new theory to describe the interior of black holes, a quantum theory of gravity.

What did Wheeler think of the importance of Penrose's 1965 theorem, the paper that in 2020 led Penrose to be the first person to be awarded a Nobel Prize for theoretical work on GR? We find the answer to this question in Box 20 of the John Archibald Wheeler Papers at the American Philosophical Society. Within it the folder "Penrose, Roger" contains a letter by W.H. McCrea soliciting a letter of evaluation of Penrose's accomplishments to decide whether Penrose should become a member of the Royal Society.[61] Wheeler sent McCrea a two-page letter on 2 December 1969, comparing Penrose with the "other outstanding men in general relativity in the 30 to 50 year age range," and ranking Penrose at the top of the field, together with Yakow Zel'dovich, Charles Misner and Kip Thorne. He elaborated: "Each of the first four stands at the top of the quartet in at least one regard: Zel'dovich in overall physical insight; Penrose in powers of mathematical analysis and the absolutely unique depth of his geometrical insight; Misner in physical originality; and Thorne in his unmatched productive power in applying general relativity to issues of lively astronomical interest." Wheeler then goes on to write in some detail of six of Penrose's accomplishments: his reformulation of GR in terms of spinor calculus, his "remaking" of the Petrov classification using the spinor formulation of GR, his treatment of asymptotic infinity and the related invention of Penrose diagrams, his development of twistor theory, his derivation of the "peeling theorem" together with Ray Sachs, and his derivation, together with Ted Newman, of the so-called NP-constants in GR. Wheeler concludes by saying that "I know no one who even approaches him in intuitive feel for the content of alternative formulations of geometry, and no one who comes close in his lectures and writings bringing such a wealth of insights to the service and inspiration of his colleagues."[62]

And yet, there is *no mention whatsoever* of Penrose's singularity theorem in Wheeler's list of Penrose's most important accomplishments, though he was clearly aware of it when he wrote that letter, having discussed it in detail in the paper cited above, just a year before. This did not change when Wheeler wrote a similar letter to D.S. Jones on 3 January 1972, which reinforces the interpretation that Wheeler did *not* think of Penrose's theorem as establishing something crucial—Wheeler had, of course, already been strongly convinced that GR needed to be replaced by a quantum theory of gravity, and so Penrose's theorem was, in his mind, likely only unnecessary grist for this mill.[63]

Let us now come back to Landsman's argument that Penrose's 1965 theorem needed an extra ingredient, soon delivered by Penrose himself, to establish the claim that GR implies that black holes necessarily form in gravitational collapse.

As we just saw from the above Cheshire cat quote from Wheeler—and in a way already from Oppenheimer and Snyder—a characteristic feature of a black hole was soon believed to be its "blackness," the feature that no light would be able to escape a black hole, even if a beam of light were to be sent outwardly from within.[64] This feature depends on the astronomical object in question having a trapped surface. Of course, as noted, the Schwarzschild, Reissner-Nordström and Kerr solutions all feature trapped surfaces, indeed even event horizons.[65] Still, in Penrose's theorem the existence of a trapped surface in the given spacetime is a premise. What Penrose's theorem has shown is that in spacetimes in which these premises hold, a singularity will arise, as first expected by Oppenheimer and Snyder for the pressureless collapsing fluidball with an exterior Schwarzschild field on the outside in particular. One might ask for even more, namely the prediction of the dynamical emergence of a trapped surface in a spacetime that does not feature it yet. This further aim is linked to the aim of proving one version or the other of Penrose's cosmic censorship hypotheses, a concept that he pioneered and worked out in the years after 1965.

In 1969, Penrose first introduced the cosmic censorship hypothesis as the question whether in situations such as gravitational collapse there "exist[s] a 'cosmic censor' who forbids the appearance of naked singularities, clothing each one in an absolute event horizon."[66] Spelling this out soon led to a proliferation of "weak" and "strong" types of cosmic censorship, both in Penrose's work and in that of his contemporaries, and onwards to today—indeed, defining, probing and attempting to prove versions of both kinds of cosmic censorship is still arguably one of the most important construction sites of mathematical relativists today.[67]

Landsman's argument is that developing the weak cosmic censorship hypothesis is what closed the gap between Penrose's theorem and his showing that not only spacetime singularities but also black holes develop in gravitational collapse: "Roughly speaking, to infer from (null) geodesic incompleteness [i.e., the singularity conclusion of Penrose's 1965 theorem] that there is a "black" object one needs weak cosmic censorship, whereas in addition a "hole" exists (as opposed to a boundary of an extendible space-time causing the incompleteness of geodesics) if strong cosmic censorship holds."[68] I will have to leave an evaluation of this, and indeed tracing the different versions of cosmic censorship, to future work; for now let us take stock and conclude.

**Conclusion**

What Einstein and Schwarzschild thought of as the "Schwarzschild singularity" and what I called the spherical singularity turned out to be only a coordinate artifact and one that hid something entirely different. Where Einstein and Schwarzschild had located the spherical singularity there is actually an event horizon, something very different from a singularity, a concept that had not been available to Einstein and Schwarzschild but that only emerged about fifty years after their first deliberations on how to interpret the Schwarzschild solution.

Einstein and Schwarzschild had thought of the center of the Schwarzschild solution as harmless: a placeholder for the star itself. But Penrose showed that for a collapsing star, even one without perfect spherical symmetry, a singularity like this will *actually* arise in nature. If GR is right, then such a singularity is not just a mathematical placeholder but something that actually exists in nature: it comes into existence as the result of collapsing matter, in the formation of a black hole. This is the primary insight of Penrose's work of 1965.

But what about Einstein's anti-singularity argument, his argument against the Hadamard catastrophe?

Well, remember that Einstein's argument went against the possibility that the spherical "singularity" could ever end up outside of a star. Oppenheimer and Snyder showed that this is possible, and thus that black holes are possible. And Finkelstein, Rindler and Penrose showed that the spherical "singularity" is indeed *not* a singularity *but* that there is an event horizon in the same spot. Would Einstein be reconciled by this? I think not. For the central singularity to appear in nature would have been *equally* catastrophic for him, now that it can't be interpreted as a placeholder for matter but must be seen as the result of collapsing matter, and thus as something that actually appears in the universe. Even worse for Einstein: Penrose showed that the appearance of such singularities is a generic feature of general relativity, something that does not only appear for a single solution like the Schwarzschild solution but that is actually a *typical* occurrence in many solutions of the Einstein field equations.

And what about Wheeler's skepticism about spacetime singularities appearing in nature? Does the existence of black holes necessitate the existence of spacetime singularities? Yes, if GR is trustworthy all the way down into a black hole. Penrose's singularity theorem shows that singularities are an inescapable consequence of GR, and that they will inescapably form in gravitational collapse as described by GR. But as Penrose would be the first to admit (and indeed did admit already in the 1965 paper): Wheeler's dream of complementing classical GR with a quantum theory of gravity that kicks in beyond the event horizon, and might stop singularities from happening, from coming into existence, remains an open and viable project.

In the course of this paper I have distinguished between eight interpretations of the presence of singularities in a given solution to the Einstein equations: the placeholder, pragmatic, barrier or black shield, indicator, problem/breakdown, transformative or black membrane interpretations, as well as barrier interpretation 2.0, and indicator interpretation 2.0. Which of these interpretations are still viable options today?

It seems to me that you can still use the placeholder interpretation exactly in the same way as it was used by Einstein and Schwarzschild: you can still use the Schwarzschild solution as a representative of the exterior gravitational field of the Sun (and indeed of black holes) and interpret the central singularity as a placeholder for a model of the Sun or the matter core of the black hole itself. In contrast, the pragmatic interpretation of the spherical singularity that Eisenstaedt attributed to most workers in the field during the 1920s is not available anymore. This is because the interpretation rests on seeing the spherical singularity in the Schwarzschild metric as an actual mathematical singularity, rather than a coordinate artifact, even though one that does not correspond to anything in nature. However, we have learned from Finkelstein that there is actually not a real singularity at $r = 2m$ but something else (and something real) instead. Finkelstein called it a unidirectional membrane, Rindler and Penrose an event horizon. Likewise the barrier or black shield interpretation of the spherical "singularity" is not available anymore, for it too rests on interpreting $r = 2m$ as a real, physical singularity. In contrast, the barrier interpretation 2.0 can still be used/endorsed: if you are unwilling to straight out identify "geodesic incompleteness" with "singular spacetime" and instead reserve the latter term for spacetimes with a semi-localizable singularity, for example a curvature singularity, then the barrier interpretation 2.0 is a way to connect geodesic incompleteness and this more demanding notion of "singularity" appropriately. But you would have to accept that not every geodesically incomplete spacetime has a semi-localizable singularity, though every spacetime with such a singularity is geodesically incomplete.

The problem/breakdown interpretation of the central singularity is also still viable, and as noted is still used by quantum gravity researchers to argue for the need of a quantum theory of gravity. Finally, the transformative or black membrane interpretation of the spherical "singularity" also still stands, but as already with Finkelstein only for *some* singularities: something that appears as a singularity in but one coordinate system may actually signify the location of an event horizon, like in the case of the spherical "singularity" of the Schwarzschild spacetime. But not *every* singularity can be reinterpreted thus. Still, it is important to note that just asking whether a singularity that you have just found in your newly discovered solution to the Einstein equations is a coordinate singularity or not is not enough: even if it is "just" a coordinate singularity, it might *also* point to something else in disguise, as witnessed by the Schwarzschild spacetime.

Finally, *both* versions of the indicator interpretation can still be used. The original indicator interpretation is still very much alive in the community working on specific exact solutions of the Einstein equations; and a singularity in a new exact solution is often interpreted as indicative of material sources. Likewise, the indicator interpretation 2.0 can still be used: discovering that a spacetime is geodesically incomplete may motivate you to look at whether you can find a semi-localizable singularity in that spacetime that could then be interpreted as the reason for the geodesic incompleteness.

One might be tempted to ask: so what is the correct or at least the best interpretation of "singularity"? I think it would be a mistake to try to give a definitive answer on this. The different interpretations presented above originated not only during different times but also among different sub-communities of relativists. They are tools fashioned to serve particular purposes, either to work on and interpret particular exact solutions or to prove and interpret general theorems that cover huge parts of the solution space of the Einstein equations. The placeholder interpretation and the original indicator interpretation originated and are particularly useful in the former context whereas the indicator interpretation 2.0, and more generally the idea of linking singular spacetimes to geodesics incompleteness, originate and are most useful in the latter context.

I think philosophers too often ask "What is the right or true interpretation of this concept or theory?" To understand scientific practice they should much more often ask: "Which interpretation is the best tool for a given purpose?"

I would like to finish with two quotes. The first is from Jerry B. Griffiths and Jiří Podolský's 2009 book *Exact Space-Times in Einstein's General Relativity*. They wrote: "It appears that it is much easier to find a new solution of Einstein's equations than it is to understand it."[69] As we have seen, it took Schwarzschild only a few months to find the Schwarzschild solution whereas it took more than half a century to fully understand what the solution has to offer; and yet it is the simplest exact solution to the vacuum Einstein equations (apart from Minkowski spacetime, one might object). So I think Griffith and Podolsky have a point.

Here is a second quote, maybe unexpectedly from Friedrich Hund. Although writing about the history of quantum mechanics (in which he played a major role), his statements apply here too: "How difficult were these advances? If we now acquaint ourselves with some of the detours, errors and psychological blocks, it is not to console ourselves with the thought that even the great physicists do not always move in a straight line, but rather to grasp in some measure how difficult it really was."[70]

**Acknowledgments**
I am very grateful for the audience of an earlier version of this paper at the Foundations seminar of Harvard's Black Hole Initiative; I am particularly grateful to Jeremy Butterfield, Erik Curiel, Jamee Elder, Sam Fletcher, Klaas Landsman, Martin Lesourd, Niels Martens, Tiffany Nichols, and John Norton. I am particularly grateful to Michel Janssen for his careful editing of the paper, to Christian Röken for reading an earlier draft of this paper extremely carefully and providing me with crucial feedback, and to Evangelia Siopi for help with the French sources, in particular for marvelous translations of Nordmann's report of the 1922 Paris conference and of the transcription of the debates at the conference. Most of all, I am grateful to Sir Roger Penrose himself, for many enlightening discussions and sharing of his memories with me. This work was funded by the European Research Council, Grant 101088528 COGY, and by the Lichtenberg Grant for Philosophy and History of Physics of the Volkswagen Foundation.

**References**

[1] Werner Israel, "Dark Stars: The Evolution of an Idea," in *Three Hundred Years of Gravitation*, eds. Stephen Hawking and Werner Israel (Cambridge, UK: Cambridge University Press, 1987), 199–276, 253.

[2] Both quotes are from the Nobel Foundation press release, "The Nobel Prize in Physics 2020," NobelPrize.org. Nobel Prize Outreach 2025, https://www.nobelprize.org/prizes/physics/2020/summary/.

[3] See "Astronomers reveal first image of the black hole at the heart of our galaxy," https://eventhorizontelescope.org/blog/astronomers-reveal-first-image-black-hole-heart-our-galaxy, and Event Horizon Telescope Collaboration, *et al.* "First Sagittarius a* Event Horizon Telescope Results. i. The Shadow of the Supermassive Black Hole in the Center of the Milky Way," *The Astrophysical Journal Letters* **930**, no. 2 (2022), L12, https://iopscience.iop.org/article/10.3847/2041-8213/ac6674.

[4] Roger Penrose, "Gravitational Collapse and Space-Time Singularities," *Physical Review Letters* **14**, no. 3 (1965), 57–59.

[5] John Archibald Wheeler, "Our Universe: The Known and the Unknown," *The American Scholar* (1968), 248–74.

[6] Klaas Landsman, "Penrose's 1965 Singularity Theorem: From Geodesic Incompleteness to Cosmic Censorship," *General Relativity and Gravitation* **54**, no. 10 (2022), 115, 3 (labelling Landsman's bullet points [a] and [b]).

[7] See, e.g., Jeffrey Crelinsten, *Einstein's Jury. The Race to Test Relativity* (Princeton: Princeton University Press, 2006) and Daniel Kennefick, *No Shadow of a Doubt: The 1919 Eclipse Expedition That Confirmed Einstein's Theory of Relativity* (Princeton: Princeton University Press, 2019).

[8] The Ricci scalar $R$ vanishes if $T_{\mu\nu}$ vanishes; for details, see, e.g., Robert M. Wald, *General Relativity* (Chicago: University of Chicago Press, 1984), 72.

[9] Einstein's calculation of Mercury's anomalous perihelion rested on previous work with Michele Besso in the context of the so-called *Entwurf* theory, the 1913 predecessor of GR; see Michel Janssen's paper in this issue for details.

[10] For more details on the astronomical parameters that Einstein had to assume in order to make this prediction see Dennis Lehmkuhl, "Literal vs. Careful Interpretations of Scientific theories: The Vacuum Approach to the Problem of Motion in General Relativity," *Philosophy of Science* **84**, no. 5 (2017), 1202–14; extended version at https://philsci-archive.pitt.edu/12461/.

[11] The correspondence between Einstein and Schwarzschild on the matter can be found in Robert Schulmann, A. J. Kox, Michel Janssen and József Illy, eds. *The Collected Papers of Albert Einstein* (Princeton: Princeton University Press, 1998), 8 (CPAE8). Schwarzschild's results were soon published as Karl Schwarzschild, "Über das Gravitationsfeld eines Massenpunktes nach der Einsteinschen Theorie," *Sitzungsberichte der Königlich Preussischen Akademie der Wissenschaften* (1916), 189–96.

[12] For details, see, e.g., Ray D'Inverno, *Introducing Einstein's Relativity* (Oxford: Oxford University Press, 1992), 189, or Sean Carroll, *Spacetime and Geometry: An Introduction to General Relativity* (Harlow: Pearson, 2013), 196.

[13] For details and context of what I call Einstein's placeholder interpretation of singularities, see Lehmkuhl, "Literal vs. Careful Interpretations" (ref. 10) and Dennis Lehmkuhl, "The Genesis of

Einstein's Work on the Problem of Motion in General Relativity," *Studies in History and Philosophy of Modern Physics* **67** (2019), 167–90.

[14] Jean Eisenstaedt, "The Early Interpretation of the Schwarzschild Solution," in *Einstein and the History of General Relativity*, eds. Don Howard and John Stachel (Boston: Birkhäuser, 1989), 213–33. See also Galina Weinstein, "Einstein's Legacy: From General Relativity to Black Hole Mysteries," especially its Chapter 2; Springer, Cham, 2024.

[15] Hermann Weyl, "Zur Gravitationstheorie," *Annalen der Physik* **359**, no. 18 (1917), 117–45; in Hermann Weyl, *Gesammelte Abhandlungen*, ed. K. Chandrasekharan (Berlin: Springer, 1968), 1: 685.

[16] Karl Schwarzschild, "Über das Gravitationsfeld einer Kugel aus inkompressibler Flüssigkeit nach der Einsteinschen Theorie," *Sitzungsberichte der Königlich Preussischen Akademie der Wissenschaften* (1916), 424–34.

[17] See Charles Nordmann, "Einstein expose et discute sa théorie," *Revue des Deux Mondes (1829-1971)* **9**, no. 1 (1922), 129–66. For a report on Einstein's Paris visit more generally, see Charles Nordmann, "Revue scientifique: Einstein à Paris," *Revue des Deux Mondes (1829-1971)* **8**, no. 4 (1922), 926–37. For detailed discussion of the political dimension of Einstein's visit to Paris and the reports of the German ambassador to the foreign service, see Siegfried Grundmann, *The Einstein Dossiers: Science and Politics—Einstein's Berlin Period with an Appendix on Einstein's FBI File* (Cham: Springer, 2005). This was the first time that a German academic (not just a physicist) had been invited to France since the end of the first world war, and it was seen as diplomatically incredibly important. Grundmann's book details how Einstein acted as an unofficial ambassador (but always watched and often supported by the German diplomatic corps) for a reintegration of German science into the international community. As to Nordmann's report, on which the following account is based, it must be noted that it was meant to be accessible to the general reader and contains only few direct quotes from Einstein; thus it must be used with caution.

[18] For a detailed argument and historical evidence of the claims made in this paragraph, see Lehmkuhl, "The Genesis of Einstein's Work on the Problem of Motion in General Relativity" (ref. 13).

[19] Arthur Stanley Eddington, *Space, Time and Gravitation: An Outline of the General Relativity Theory* (Cambridge: Cambridge University Press, 1920), 98.

[20] I owe this list to Jean Eisenstaedt, "La relativité générale à l'étiage: 1925–1955," *Archive for History of Exact Sciences* **35** (1986), 115–85.

[21] Arthur Stanley Eddington, *The Mathematical Theory of Relativity* (Cambridge: Cambridge University Press, 1923), 165.

[22] It can be found in various chapters of Jerry B. Griffiths and Jiří Podolský, *Exact Space-Times in Einstein's General Relativity* (Cambridge: Cambridge University Press, 2009).

[23] For Eddington's role during the further development of the theory of the interior of stars, see Israel, "Dark stars" (ref. 1), especially Sections 7.3 and 7.4.

[24] The limit is today called the Chandrasekhar limit. For details of the contemporary discussion, including Eddington's opposition to Chandrasekhar's results, see Israel, "Dark stars" (ref. 1).

[25] J. Robert Oppenheimer and Hartland Snyder, "On Continued Gravitational Contraction," *Physical Review* **56**, no. 5 (1939), 455–59.

[26] Israel, "Dark stars" (ref. 1), 226.

[27] Oppenheimer and Snyder were relying on a new exact solution to the full Einstein equations by

Richard C. Tolman, "Static Solutions of Einstein's Field Equations for Spheres of Fluid," *Physical Review* **55**, no. 4 (1939), 364–73.

[28] Oppenheimer and Snyder, "Continued Gravitational Contraction" (ref. 25), 459.

[29] In a joint paper with Jeroen van Dongen, I will analyze in detail Einstein's skepticism about these assumptions and the context of a 1939 paper by Einstein that could be read as opposing Oppenheimer and Synder: Albert Einstein, "On a Stationary System with Spherical Symmetry Consisting of Many Gravitating Masses," *Annals of Mathematics* **40**, no. 4 (1939), 922–36. For the current paper, the most important thing is that in 1965 Roger Penrose would show that the major results of Oppenheimer and Snyder, that of continuing collapse and that of a formation of a central singularity, do not depend on these assumptions (see below).

[30] J. Barclay Adams, B. Kent Harrison, Louis T. Klauder, Jr., Raymond Mjolsness, Masami Wakano, John Archibald Wheeler, and Raymond Willey, "Some Implications of General Relativity for the Structure and Evolution of the Universe" in *La Structure et l'évolution de l'univers: rapports et discussions* (Brussels: Institut international de physique Solvay, 1958), pp. 125–41, signed only by Harrison, Wakano, and Wheeler.

[31] For further context and discussion, see Israel, "Dark stars" (ref. 1), 230 and Adams, Harrison, Klauder, Jr., Mjolsness, Wakano, Wheeler and Willey, "Some Implications of General Relativity for the Structure and Evolution of the Universe," (ref. 30).

[32] David Finkelstein, "Past-Future Asymmetry of the Gravitational Field of a Point Particle," *Physical Review* **110**, no. 4 (1958), 965–67, 965. Lemaître had earlier seen (as clearly as Finkelstein) that $r = 2m$ is, as he put it, but a "fictitious singularity." See section 11 in George Lemaître, "L'univers en expansion," *Annales de la Société Scientifique de Bruxelles* **53** (1933), 51.

[33] Penrose, "Gravitational Collapse and Space-Time Singularities" (ref. 4), 57.

[34] Evgeny Mikhailovich Lifshitz and Isaak Markovich Khalatnikov, "Investigations in Relativistic Cosmology," *Advances in Physics* **12**, no. 46 (1963), 185–249. See Israel, "Dark stars" (ref. 1), 251 for references regarding how this result was later reconciled with Penrose's 1965 theorem.

[35] Maarten Schmidt, "3C 273: A Star-like Object with Large Red-shift." *Nature* **197**, no. 4872 (1963), 1040.

[36] Fred Hoyle and William A. Fowler, "Nature of Strong Radio Sources," *Nature* **197**, no. 4867 (1963), 533–35, 535.

[37] Quoted in Israel, "Dark Stars" (ref. 1), 244; I assume that Israel quotes from the invitation that he himself received at the time.

[38] Interview of Roger Penrose by Alan Lightman on January 24, 1989, Niels Bohr Library & Archives, American Institute of Physics, College Park, MD USA, https://www.aip.org/history-programs/niels-bohr-library/oral-histories/34322.

[39] Roger Penrose, "Gravitational Collapse and Space-Time Singularities" (ref. 4), 57.

[40] Penrose's theorem is stated in this manner in José María Martín Senovilla and David Garfinkle, "The 1965 Penrose Singularity Theorem," *Classical and Quantum Gravity* **32**, no. 12 (2015), 124008. Senovilla and Garfinkle argue that this way of writing it also gives the structure of every subsequent singularity theorem including the later singularity theorems by Hawking and Hawking and Penrose jointly, for which Penrose's 1965 theorem served as a blueprint. Indeed, the later singularity theorems can be put into the form below, but with a different curvature condition, a

different causality condition, and a different initial or boundary condition implying that the spacetime in question is geodesically incomplete.

[41] As pointed out in Landsman, "Penrose's 1965 Singularity Theorem" (ref. 6, 115), Lemaître was the first (in 1933; see ref. 32) to use an energy condition instead of a specific energy-momentum tensor, followed by Arthur Komar, "Necessity of Singularities in the Solution of the Field Equations of General Relativity," *Physical Review* **104**, no. 2 (1956), 544–46. However, this only became common usage after Penrose's 1965 paper (ref. 4).

[42] Landsman, "Penrose's 1965 Singularity Theorem" (ref. 6), 115, argues that this specific way of defining a Cauchy surface "seems to have originated with Penrose", and that he thus "should be included on the list of architects of global hyperbolicity in GR" with Leray, Choquet-Bruhat and Geroch.

[43] Penrose, "Gravitational Collapse and Space-Time Singularities" (ref. 4), 58.

[44] Roy P. Kerr, "Gravitational Field of a Spinning Mass as an Example of Algebraically Special Metrics," *Physical Review Letters* **11**, no. 5 (1963), 237–38. Kerr had given a talk on his solution at the First Texas Conference.

[45] Penrose, "Gravitational Collapse and Space-Time Singularities" (ref. 4), 58.

[46] Kip S. Thorne, *Black Holes & Time Warps: Einstein's Outrageous Legacy* (New York: Norton, 1994), 187.

[47] See p. 415 for Penrose's editorial comment of Volume 1 of his Collected Works, in front of the reprinting of the Royamount paper in Roger Penrose, *Collected Works* (Oxford: Oxford University Press, 2010), 1:415–20 (Chapter 14).

[48] The remark that the larger portion of the paper originated during his stay in Princeton and Syracruse in 1960–61 appears in the first footnote on the first page of the paper: Roger Penrose, "A Remarkable Property of Plane Waves in General Relativity," *Reviews of Modern Physics.* **37**, no. 1 (1965), 215-220.

[49] See Dennis Lehmkuhl, Christian Röken and Juliusz Doboszewski, "On Penrose's Analogy between Curved Spacetime Regions and Optical Lenses," *Philosophy of Physics* **2**, no. 1 (2024), 4, 1–40, https://doi.org/10.31389/pop.54.

[50] Ezra Newman and Roger Penrose, "An approach to Gravitational Radiation by a Method of Spin Coefficients," *Journal of Mathematical Physics* **3**, no. 3 (1962), 566–78.

[51] Roger Penrose, "Zero Rest-mass Fields Including Gravitation: Asymptotic Behaviour," *Proceedings of the Royal Society of London. Series A. Mathematical and Physical Sciences* **284**, no. 1397 (1965), 159–203.

[52] Charles W. Misner, "The Flatter Regions of Newman, Unti, and Tamburino's Generalized Schwarzschild Space," *Journal of Mathematical Physics* **4** no, 7, (1963), 924–37. See Landsman, "Penrose's 1965 Singularity Theorem" (ref. 6), for further authors who studied geodesic incompleteness in Lorentzian manifolds before Penrose in 1965, though often without constructing an explicit link to how singular spacetimes are to be defined.

[53] Penrose, "Gravitational Collapse and Space-Time Singularities" (ref. 4), 58.

[54] P. 257 of Roger Penrose, "Gravitational Collapse: The Role of General Relativity", Rivista del Nuovo Cimento, Serie 1, Vol 1, Numero Speciale, 257.

[55] Erik Curiel has convincingly argued that geodesic incompleteness is not a sufficient condition for there to be a semi-localizable singularity. However, that is consistent with how I introduced

the "indicator" term above: it's not a sufficient condition, but it is a reason to look for something like that. See Erik Curiel, "The Analysis of Singular Spacetimes," *Philosophy of Science* **66**, no. S3, (1999), S119–S145.

[56] Stephen Hawking, "Properties of Expanding Universes," (Ph.D. diss., University of Cambridge, 1966), section 4.1, p. 96. I was first alerted to the fact that the move to defining "singular spacetime" thus originated with Hawking's Ph.D. thesis (which built directly on Penrose's work) by Klaas Landsman in April 2021.

[57] John Archibald Wheeler and Kenneth W. Ford, *Geons, Black Holes and Quantum Foam: A Life in Physics* (New York: Norton, 1998), Chapter 13, p. 296.

[58] Wheeler, "Our Universe" (ref. 5), 8–9.

[59] For more sources on the subsequent use of "black hole" in the years immediately after 1968 see Klaas Landsman, "Singularities, Black Holes, and Cosmic Censorship: A Tribute to Roger Penrose," *Foundations of Physics* **51**, no. 2 (2021), 41: 1–38, 30–31, "Appendix: A Potted Early History of "black hole" (by Erik Curiel)."

[60] Wheeler, "Our Universe" (ref. 5), 11.

[61] W.H. McCrea to John Archibald Wheeler, November 17, 1969, "Penrose, Roger" folder, Box 20, John Archibald Wheeler Papers, American Philosophical Society, Philadelphia, PA.

[62] John Archibald Wheeler to W.H. McCrea, December 2, 1969, "Penrose, Roger" folder, (ref. 61).

[63] John Archibald Wheeler to D.S. Jones, January 3, 1972, "Penrose, Roger" folder, (ref. 61).

[64] I shy away from calling this the defining rather than a characteristic feature of a black hole, for there are different operative definitions of "black hole" even today. See, e.g., Erik Curiel, "The Many Definitions of a Black Hole," *Nature Astronomy* **3**, no. 1 (2019), 27–34.

[65] Every spacetime that contains an event horizon contains a trapped surface but not vice versa.

[66] Roger Penrose, "Gravitational Collapse: The Role of General Relativity" (ref. 54), 274.

[67] See, e.g., Frank J. Tipler, Christopher J. S. Clarke and George F. R. Ellis, "Singularities and Horizons," *General Relativity and Gravitation II* **2** (1980), 97–206; John Earman, *Bangs, Crunches, Whimpers, and Shrieks: Singularities and Acausalities in Relativistic Spacetimes* (Oxford: Oxford University Press, 1995), Chapter 3; Landsman, "Penrose's 1965 singularity theorem" (ref. 6), section 5 and 6; Landsman, "Tribute to Roger Penrose" (ref. 59); Yen Chin Ong, "Space–time Singularities and Cosmic Censorship Conjecture: A Review with Some Thoughts," *International Journal of Modern Physics. A* **35**, no. 14 (2020), 2030007; as well as references cited in these sources.

[68] Landsman, "Penrose's 1965 singularity theorem" (ref. 6), 1.

[69] Griffiths and Podolský, *Exact Space-Times* (ref. 22), XV.

[70] Friedrich Hund, "Irrwege und Hemmungen beim Werden der Quantentheorie," in *Quanten und Felder,* ed. H.P. Dürr, (F. Vieweg: Braunschweig, 1971), 1-10, 1.